\documentclass[conference]{IEEEtran}
\IEEEoverridecommandlockouts
\usepackage{cite}
\usepackage{graphicx}  
\graphicspath{{./Pics/}}
\usepackage{amsmath,amssymb,amsfonts}
\usepackage{pseudocode}
\usepackage{multirow}
\usepackage{balance}
\usepackage{multicol, lipsum}

\usepackage{textcomp}
\usepackage{balance}
\usepackage{xcolor}
\usepackage{algorithm,algpseudocode}
\def\BibTeX{{\rm B\kern-.05em{\sc i\kern-.025em b}\kern-.08em
    T\kern-.1667em\lower.7ex\hbox{E}\kern-.125emX}}

\algnewcommand\algorithmicforeach{\textbf{for each}}
\algdef{S}[FOR]{ForEach}[1]{\algorithmicforeach\ #1\ \algorithmicdo}

\title{\huge Handover Management through Reconfigurable Intelligent Surfaces for VLC under Blockage Conditions\vspace{-4mm}}

\author{\IEEEauthorblockA{\normalsize{Kapila W. S. Palitharathna}\IEEEauthorrefmark{1}, \normalsize{}{Anna Maria Vegni}\IEEEauthorrefmark{2}, \normalsize{Panagiotis D. Diamantoulakis}\IEEEauthorrefmark{3}, \normalsize{Himal A. Suraweera}\IEEEauthorrefmark{4}, \normalsize{Ioannis Krikidis}\IEEEauthorrefmark{1}}\IEEEauthorblockA{\IEEEauthorrefmark{1}Department of Electrical and Computer Engineering, University of Cyprus, Nicosia, Cyprus\\}\IEEEauthorblockA{\IEEEauthorrefmark{2}Department of Industrial, Electronic and Mechanical Engineering, Roma Tre University, Rome, Italy\\}\IEEEauthorblockA{\IEEEauthorrefmark{3}Department of Applied Informatics, University of Macedonia, Thessaloniki, Greece\\}\IEEEauthorblockA{\IEEEauthorrefmark{4}Department of Electrical and Electronic Engineering, University of Peradeniya, Peradeniya, Sri Lanka\\}\vspace{-10mm}
}

\begin{document}

\maketitle

\begin{abstract}
In this paper, we consider an indoor visible light communication (VLC) system with multiple “white” light emitting diodes serving to form overlapping wireless communication cells. In order to maintain seamless connectivity to mobile users, a handover procedure should be implemented. In particular, practical conditions such as blockages due to obstacles inside the room environment and the mobility of users can affect direct VLC connectivity. The use of reconfigurable intelligent surfaces (RISs) in optical wireless systems allows to exploit non-direct connectivity links, thus providing efficient communication links. In this paper, we present a proactive handover mechanism that exploits the presence of a RIS, in order to redirect the communication links in case of blockages. The proposed approach has been implemented both in hard and soft modes and assessed in terms of achievable data rate and handover latency for a user walking in a given reference room at different user speeds and blockage conditions. Our presented results and comparisons with conventional handover methods (\textit{i.e.}, without RIS) are helpful in showing the superiority of the presented algorithm.
\end{abstract}
%\begin{IEEEkeywords}
%Visible light communication, intelligent reflecting surface, handover management, blockage.
%\end{IEEEkeywords}
%\vspace{-2mm}
\section{Introduction}
\vspace{-1mm}
%TO DO: \\
%- consider different IRS-based architectures (e.g., posed on the Tx/Rx?) \\
%- start defining the system model\\

 %   \begin{itemize}
  %  \item Introduction to VLC 
   %  and IRS for handover management
    %\item Discussion on mobility issues
    %\item Short overview of handover solutions in VLC networks
    %\item How to define blockages affecting LoS connectivity? \figurename~\ref{fig:room2} depicts an example of blockages occurring in an indoor environment.
    %\item Main goals of the paper i.e., (i) definition of blockage probability, (ii) handover algorithm driven by blockage estimation through ML techniques, (iii) simulation results in pure VLC and hibrid RF+VLC networks.
    %\item organization of the paper
    %\end{itemize}

In next generation wireless networks, the use of alternative technology bearers than radio frequency (RF) spectrum is strongly needed by a set of RF drawbacks and limitations arisen so far. Due to the huge amount of mobile devices, the RF spectrum is suffering and cannot guarantee enough bandwidth to support such a large demand of data traffic, thus causing the well-known ``spectrum crunch problem''~\cite{requirements6G}. %is indeed one of the main issues that moved the research community to investigate other frequency bands, specifically moving towards high-frequency domain.

Visible light communication (VLC) is a candidate technology for beyond 5G (B5G) networks, thanks to its high frequency bandwidth that provides high data rates by modulating the optical intensity generated by light emitting diode (LED) based luminaires~\cite{vlc6g}. VLC is the unique technology that allows simultaneously data transmission and illumination, thus representing a green technology bearer. Furthermore, due to the high directivity of line-of-sight (LoS) links, security can be guaranteed for a point-to-point data transmission. However, this condition strongly limits data propagation in case of mobile users, since the alignment between a LED and a photodetector (PD) should be maintained for effective data transmission. 
It follows that any blockage laying between a LED and the PD, or shadow dimming the LED or the PD, can affect the LoS condition and then reduce the quality-of-service (QoS) level of the data transmission~\cite{Anand_2021}. 

Handover management techniques should be considered in order to guarantee an ubiquitous and seamless LoS connectivity. 
For this aim, overlapping lighting cells is a condition necessary to manage handover and switch LoS connectivity from one LED to another or several other LEDs. However, this requires that multiple LEDs should populate the reference environment, and that is a condition not always achievable, especially in large rooms or office. Furthermore, a handover from a serving LED to a candidate one is valid since still considering LoS condition, which cannot be relaxed~\cite{vegni_handover, msn22}. Extending the assumption to non LoS (NLoS) components due to multiple reflections of the wireless optical signal does not provide enough power at the receiver. 

The use of reconfigurable intelligent surfaces (RISs) results as a benefit for optical wireless handover management.
The RIS concept has been arisen from RF technology, allowing to control the wireless channel by manipulating the impinging RF signal. In case of optical wireless systems, RISs can be exploited in indoor environments to steer the reflected signal toward a desired receiver. 
%
%However, multiple reflections can be effective in case of a wireless signal impinging on an Intelligent Reflecting Surface (IRS), which allows to steer the reflected optical signal toward a desired receiver. The IRS concept has been arisen from RF technology, allowing to control the wireless channel by manipulating the impinging RF signal. In this case, an IRS is a two-dimensional planar surface composed of periodical artificial atoms (a.k.a meta-atoms), through which the impinging electromagnetic wave can produce controllable induction current patterns. 
%
%In case of optical domain, 
Specifically in the visible frequency range, RIS can be physically implemented by mirror-based array or metasurfaces~\cite{Aboagye,survey_irs_vlc}. The RIS device can be deployed both at the transmitter/receiver side and in the wireless environment and different positions of the RIS allow to achieve different features~\cite{marco_survey}. 
%For instance, if the IRS is deployed at the transmitter it accomplishes beam steering features, while if it is placed on the PD, it actively expands the field-of-view (FoV). 

In this paper, we are interested in the RIS deployment for handover management in an indoor environment. The RIS acts as a reflecting surface that redirects the impinging wireless optical signals to the PD. If direct LoS connectivity 
%link from the LED to the PD 
is not supported due to blockages, the RIS can act as a bridge, where the optical signal from the LED is steered to the RIS and then redirected to the PD, providing seamless connectivity to the final user. 

% For Conference paper this part is not required

%This paper is organized as follows. Section~\ref{sec:rel_work} presents an overview of recent works dealing with handover management in VLC networks. 
%Section~\ref{sec:model} introduces the reference system model, while in 
%Section~\ref{sec:handover} we describe the proposed proactive IRS-assisted handover technique. In Section~\ref{sec:results}, the assessment of the proposed approach is shown in terms of achieved handover data rates and latency, in case of both hard and soft modes. Finally, conclusions are drawn at the end of the paper. 

\noindent \textbf{Related Work:}
Handover mechanisms in VLC networks have been largely investigated by the research community, mainly for indoor systems where VLC access points coexist in conjunction with RF technology. 
In~\cite{Wang}, Wang and Haas investigate a dynamic load balancing technique for hybrid VLC/RF systems, that is able to  mitigate unwanted and unnecessary handovers, but forcing mobile users to be served by RF access points.  In~\cite{Sathisha}, an experimental demonstration of a real-time RF/VLC data streaming system is presented. Handover occurs based on signal strength measurement, and  RF technology is exploited mainly for mobile users, while almost-static ones are preferably connected to narrow beam VLC coverage.  

Focusing on pure VLC systems, handover solutions become more complex since coverage is not always guaranteed and performance are limited to the availability of LoS connectivity links~\cite{vegni_handover}. Dinc \textit{et al.}~\cite{Dinc} investigate  soft handover methods for VLC, as well Demir \textit{et al.}~\cite{Demir} present a coordinated multipoint transmission technique in case of soft handover. Vegni and Diamantoulakis~\cite{msn22} investigate  a proactive handover mechanism for mitigation of blockages. The proposed approach allows handovers  trigged according to the presence of blockages. Both soft and hard handover occurrences are assumed, showing the best performance are achieved in case of soft handover. 
However, in all previous works it is assumed the existence of overlapping lighting cells that guarantee the handover procedure, but can cause interference. 

\noindent \textbf{Motivation and Contribution:}
In such scenario, the use of RISs for handover management can represent a suitable and effective solution. However, for the best of authors’ knowledge, no contribution exists yet addressing this topic for pure VLC scenario. The idea is similar to the work in~\cite{Jiao}, but it is investigated for mmWave systems. It was observed that, under various channel blockage conditions, it is possible to reduce the handover overhead by jointly adjusting beamformers and controlling the RIS.

Leveraging on the idea of using RIS for handover management in VLC systems under blockage conditions, this paper represents an improved version of a previous work in~\cite{msn22}, where RISs are introduced for enhanced performance.
We present an enhanced system comprised of a RIS device able to assist the handover procedure. In our vision, RIS acts as a relay node that redirects the optical intensity toward the PD, in case of blockages incurring between the transmitting LED and the PD. Differently from~\cite{msn22}, where handovers occur only assuming that overlapping wireless signals are available in each position of the environment thus strongly limiting the effectiveness of handover management, in this paper using RIS it is possible to guarantee a real seamless connectivity to the final user. Indeed, scattered and reflected signals are exploited and directed from the RIS to the PD, even if no overlapping lighting cells are available. This allows to reduce the number of LEDs, with a reduction of energy and deployment costs. 

\section{System model}\label{sec:model}
Let us consider an indoor environment where a set of $N$ LEDs that act as access points (APs) are randomly deployed on the ceiling while guaranteeing a sufficient level of illuminance and data rate constraints all over the room. The user is walking randomly while connected to his own mobile device, equipped with a PD to receive information from APs. The LED APs are assigned with an identification code that is represented by a direct-sequence optical code division multiple access (DS-OCDMA)~\cite{Hammouda:18}. Moreover, the communication is assisted by a RIS mounted on the wall, which includes $M$ number of mirror elements each having a height of $h_m$~[m] and a width of $w_m$~[m]. Each mirror element can be controlled using micro motors and the yaw and roll angle of the $j$-th mirror element (\textit{i.e.}, $\alpha_j$ and $\beta_j$) can be controlled to focus light signals on the receiver resulting in a better performance~\cite{survey_irs_vlc,Aboagye}. The mid-point of $j$-th RIS element is $(x_j,y_j)$ on the RIS plane.

    \begin{figure}[t]
    \begin{center}
\includegraphics[width=\columnwidth]{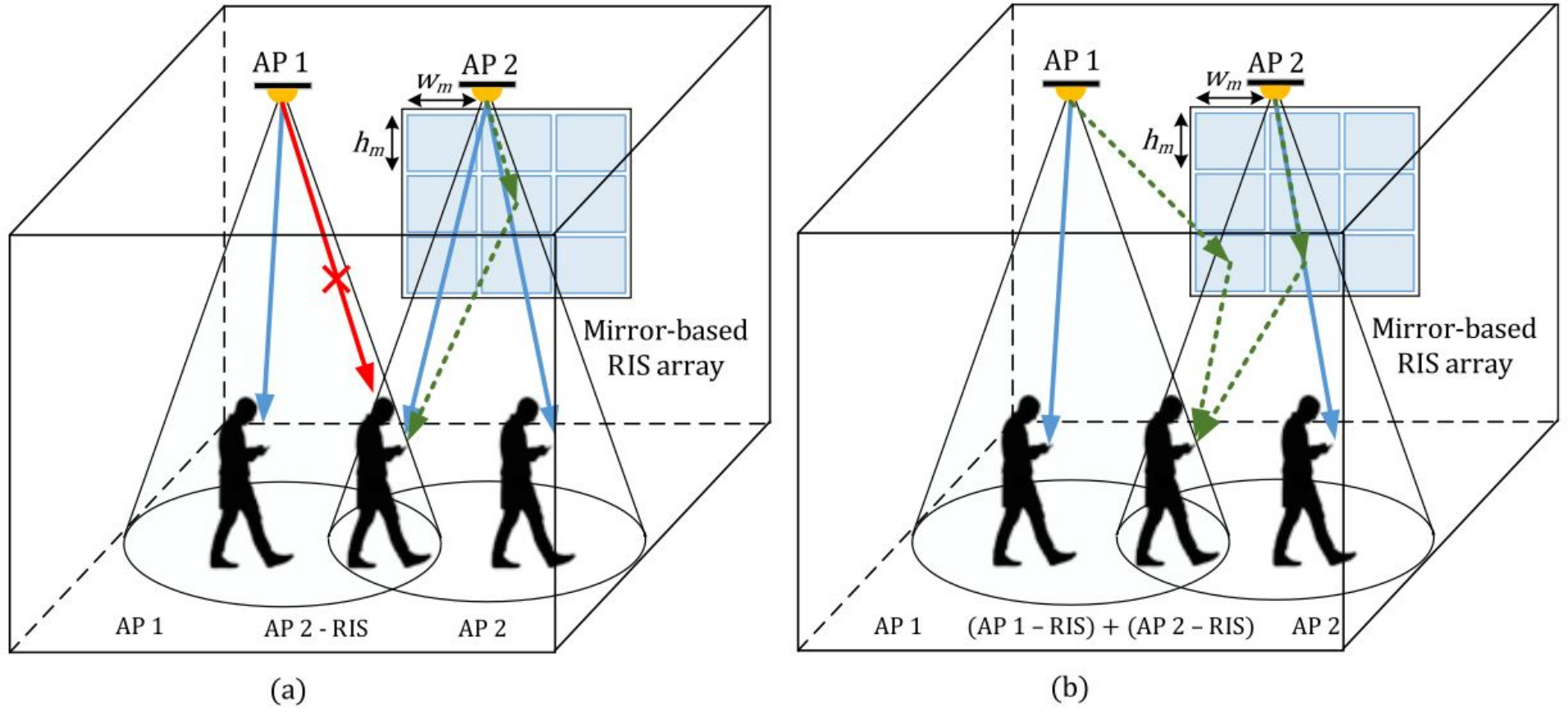}
\vspace{-6mm}
    \caption{Schematic of (\textit{a}) hard and (\textit{b}) soft handovers, for a mobile user in an overlapping lighting cells indoor scenario with the help of RIS mounted on a wall in the reference scenario.}
    \label{fig:room2}
    \end{center}
        \vspace{-8mm}
    \end{figure}
    
Overlapping light coverage may occur in this setup. Thanks to the orthogonality of the codes, we assume that the $N$ APs do not cause interference with each other. In a given position, the receiver is then able to collect the power levels from different APs by considering all the possible sequences of a given spreading factor. Specifically, the APs transmit orthogonal sequences by employing Walsh-Hadamard codes so as to grant orthogonality. In this context, we consider the special requirements for VLC transmission, particularly the condition that the transmitted symbols should be real and positive. We then adopt a  VLC-adapted version of the DS-CDMA, in which a fixed DC offset is added to convert the bipolar signal into a unipolar signal before modulating the light source.

Blockages can present due to shadows along the LoS connectivity link from the LED to the PD. To model the blockage due to humans inside the room, we use a more realistic model based on the mobility of users and a geometrical approach to determine the blockage as in~\cite{kapila_2023} and~\cite{kapila_Mass_2023}. According to this model, it is assumed that $N_B$ number of humans walk inside the room with random starting points, ending points, and speeds. While mobility, each of them can create a blockage in the LoS link from the $i$-th LED to the receiver. 

The complete optical channel gain from the $i$-th LED to the receiver can be written as~\cite{IRS_mag}
     \vspace{-1mm}
	\begin{align}\label{ch3_ch1}
	h_{i} = I_{i}h_{i}^{\text{LoS}}+\sum_{j=1}^{M}h_{i,j}(\alpha_j,\beta_j),
	\end{align}
where $I_{i}=\prod_{b=1}^{N_B}I_{i,b}$ is the blockage from the $i$-th LED to the receiver, $h_{i}^{\text{LoS}}$ is the LoS channel gain without blockage, and $h_{i,j}(\alpha_j,\beta_j)$ is the non-LoS channel gain due to the $j$-th RIS element, and $N_B$ is the number of blockages. $I_{i,b}$ is the blockage along the link from the $i$-th LED to the receiver due to the presence of the $b$-th human inside the room. In case of any blockage, then $I_{i}=0$, and the LoS channel gain will be null. If there is no blockage, $I_{i}=1$ and LoS channel gain will exist. $h_{i}^{\text{LoS}}$ is evaluated with the help of the well-known Lambertian radian pattern equation~\cite{kapila_2023}. The RIS channel gain $h_{i,j}(\alpha_j,\beta_j)$ for a given $(\alpha_j, \beta_j)$ can be modeled with the light reflection properties of a mirror and geometry similar to~\cite{nethmi_2023} and~\cite{kapila_Mass_2023}.

The exact capacity of VLC is not known. Hence, we use an achievable rate for the VLC link, which is expressed as~\cite{Lapidoth}
\vspace{-2mm}
	\begin{align}\label{R1}
	R = B\log_2\left(1+\frac{e}{2\pi}\eta\right),
	\end{align}
where $B$~[Hz] is the bandwidth of the signal, and $\eta$ is the signal-to-noise ratio (SNR). In the case of overlapping cells, the receiver combines signals coming from multiple $L_0$ LEDs (\textit{i.e.}, $L_0\ge 2$), which are transmitting the same information, and hence, the achievable data rate, $R_0$ can be written by replacing the parameter $\eta$ in Eq.~\eqref{R1} with $\sum_{i=1}^{L_0}\eta_i$ where $\eta_i$ is the SNR for the signal received from the $i$-th AP~\cite{msn22}.

%    \begin{figure}[t]
%   \begin{center}
%    \includegraphics[width=0.55\columnwidth]{Pics/room2.pdf}
%    \caption{Example of blockages in a VLC indoor network due to shadow, affecting LoS connectivity.  }
%    \label{fig:room1}
%   \end{center}
%    \vspace{-4mm}
%    \end{figure}

\section{Proactive Handover Mechanism}\label{sec:handover}
We introduce a proactive handover mechanism, with the aid of RIS, under the LoS link blockage. First, the method of identifying the blockage is introduced. Let $\xi_i$ be the blockage degree from the $i$-th LED, defined as a value in the range $[0,1]$ that represents the extent to which the channel blockage affects the channel quality. If $\xi_i= 0$, no blockage occurs in the $i$-th LoS channel, while $\xi_i = 1$ stands for a complete blockage. More in details, from~\cite{msn22} 
the blockage probability $p_{b,i}$ of the $i$-th LED is defined as $ p_{b,i} = \Pr[I_{elec,i} < \mathcal{I}]$ where $\mathcal{I}$ is a given threshold and $I_{elec, i}$ is the electric current obtained from the gathered photons at the PD, expressed as~\cite{Wu}

    \begin{equation}
    \vspace{-1mm}
    I_{elec,i} = (1-\xi_i) r h^{LoS}_{i} P_{i}, \label{eq:I_elec}
    \end{equation}
with $r$~[A/W] as the PD responsivity, $P_{i}$~[Watt] is the transmitted optical power from the $i$-th LED, and $h^{LoS}_{i}$ are the channel coefficients of LoS link from 
the $i$-th LED to the receiver. From Eq.~\eqref{eq:I_elec} 
we can easily approximate the blockage probability as follows: 
    \begin{equation}
    \vspace{-2mm}
    p_{b,i} = 
    \begin{cases}
    1 & {\rm{if}} \;\; \xi_i \geq 0.5\\
    0 & \text{otherwise}\\
\end{cases}\label{eq:p_blockage}
    \end{equation}

The handover decision is taken according to the blockage probability and availability of the RIS. At a given receiver position $P_k = (x_k,y_k)$, the blockage probability $p_{b,i}$ is calculated for the $i$-th LED. In the case of overlapping cell environments, if all the LEDs are in blockage, \textit{i.e.}, $\xi_i=1$, $\forall i$, the angles of all RIS elements are set to focus light signals coming from the closest LED to the receiver that ensures no connectivity hole. Hence with the use of RIS, the probability of having a connectivity hole is very low in a large area of the room. 

On the other hand, we compute the set of overlapping LEDs that are not in blockage for the considered position, $\mathcal{L}$. If the cardinality of $\mathcal{L}$ is equal to $1$, a hard handover will occur, while it is a soft handover in the case of $\mathcal{L}\ge 2$. In particular, the hard handover will take place when there is a link break to the serving cell, and then a link is created to a candidate cell ensuring a connectivity is established only with one AP. Moreover, the RIS angles are set in such a way that all the elements focus light signal coming from the connected AP to the receiver. For this scenario, RIS angles can be found with a geometrical method presented in~\cite{Qian_2021} once the connected AP location and the receiver location are known. It can be noted that when an RIS is used, the probability of handover becomes lower since the RIS can support the existing link even under blockage conditions. At the same time, the use of RIS increases the data rate due to high channel gain. A soft handover occurs when connectivity with more than one AP is possible. It allows the receiver to connect with more than one AP at the same time, maintaining connectivity with both the serving cell and the candidate cell. 

\figurename~\ref{fig:room2} depicts both hard and soft handovers with the assistance of RISs. In general, the data rate is higher in the case of soft handover, since the receiver is connected to several LEDs. In the case of soft handover, the RIS elements support several links and the RIS element assignment needs to be done carefully to further improve the performance. Due to the complexity of traditional RIS assignment algorithms, we use a training-based machine learning approach as explained in Subsection~\ref{subsec:ANN}. Algorithm~\ref{alg:a1} presents the key steps involved in the handover mechanism. 

\begin{figure}[t]
\vspace{-2mm}
\begin{algorithm}[H]
	\caption{\!\!: Proactive Handover Using RIS}
	\begin{algorithmic}[1]
	    \State{\textbf{Inputs:} $N$, $M$, $(x_i,y_i)$, $(x_j,y_j)$}
	    \State{\textbf{Outputs:} $R_{h/s}$, $\delta_{h/s}$, $N_{h/s}$, $(\alpha_j,\beta_j)$}
		\ForEach{$P_k = (x_k,y_k)$} 
            \State{Compute $I_{elec,i}$ using Eq.~\eqref{eq:I_elec}}
            \State{Compute $p_{b,i}$ using Eq.~\eqref{eq:p_blockage}}

            \If{$\xi_i=0, \forall i$}
                \State{Assign all RIS elements closest LED}
                \State{$R_h = R$   \Comment{$R$ calculated only with RIS channels}}
                \State{$N_h = N_h+1$ \Comment{Hard handover executions}}
            \Else{}
            \State{Compute set of LEDs not in blockage ($\mathcal{L}$)}
                \If{$|\mathcal{L}|=1$}       
                    \State{Assign all RIS elements to the selected LED}
                    \State{$R_h = R$  \Comment{Hard handover rate with RIS}}
                    \State{Compute $\delta_h$ \Comment{Hard handover latency}}
                    \State{$N_h = N_h+1$ \Comment{Hard handover executions}}
                \Else
                    \State{Assign each RIS elements among LEDs in the set $\mathcal{L}$ using the ANN architecture}
                    \State{$R_s = R_0$ \Comment{Soft handover with RIS}}
                    \State{Compute $\delta_s$ \Comment{Soft handover latency}}
                    \State{$N_s = N_s+1$ \Comment{Soft handover executions}}
                \EndIf
            \EndIf
        \EndFor
	\end{algorithmic} 
	\label{alg:a1}
\end{algorithm}
\vspace{-8mm}
\end{figure}

\subsection{ANN-based RIS Assignment}\label{subsec:ANN}
The RIS element assignment to different APs is crucial for achieving better performance, especially in the soft handover case. We use a well-trained lightweight artificial neural network (ANN) for this purpose. The use of a lightweight ANN is motivated due to its ability to learn blockage patterns induced by human mobility and the low complexity of the design that avoids additional latency to handover when RISs are used.

In this ANN architecture, the estimated blockage values, $\xi_i$, $\forall i$, and user position are used as inputs to the system while the outputs are RIS association values, $X_j$ for $j=(1,2,\ldots,M)$. The value of $X_j$ is selected from the set that includes the AP number $\{1,2,\ldots,N\}$. The ANN includes several layers as described below. First, the input is fed into a layer with $N+2$ number of nodes. It is followed by three fully connected layers with $N_1$, $N_2$, and $N_3$ number of nodes and the activation function is the rectified linear unit (ReLU). Finally, a softmax layer is used to obtain the $M$ number of RIS associations. This model is trained with $10^7$ blockage instances and corresponding RIS association vectors obtained from a brute-force simulation. 

The presented RIS-assisted handover management technique is compared with the existing handover management technique presented in~\cite{msn22} in which no RIS is used. In this method, connectivity holes can present when all links are blocked. When there is only one AP that is not in the blockage, a hard handover will occur while a soft handover will take place in the case of multiple APs are not blocked. 

\section{Numerical Results}
\label{sec:results}
Numerical results are presented to illustrate the superior performance gains of the presented RIS-assisted handover mechanism under different system and channel parameters. A room dimension of $5 \text{m}\times 5 \text{m}\times 3 \text{m}$ is assumed. The APs are assumed to be randomly placed on the ceiling. Further, simulation parameters are selected similarly to~\cite{msn22}.  

\begin{figure}[t]
    \begin{center}
    \includegraphics[width=0.8\columnwidth]{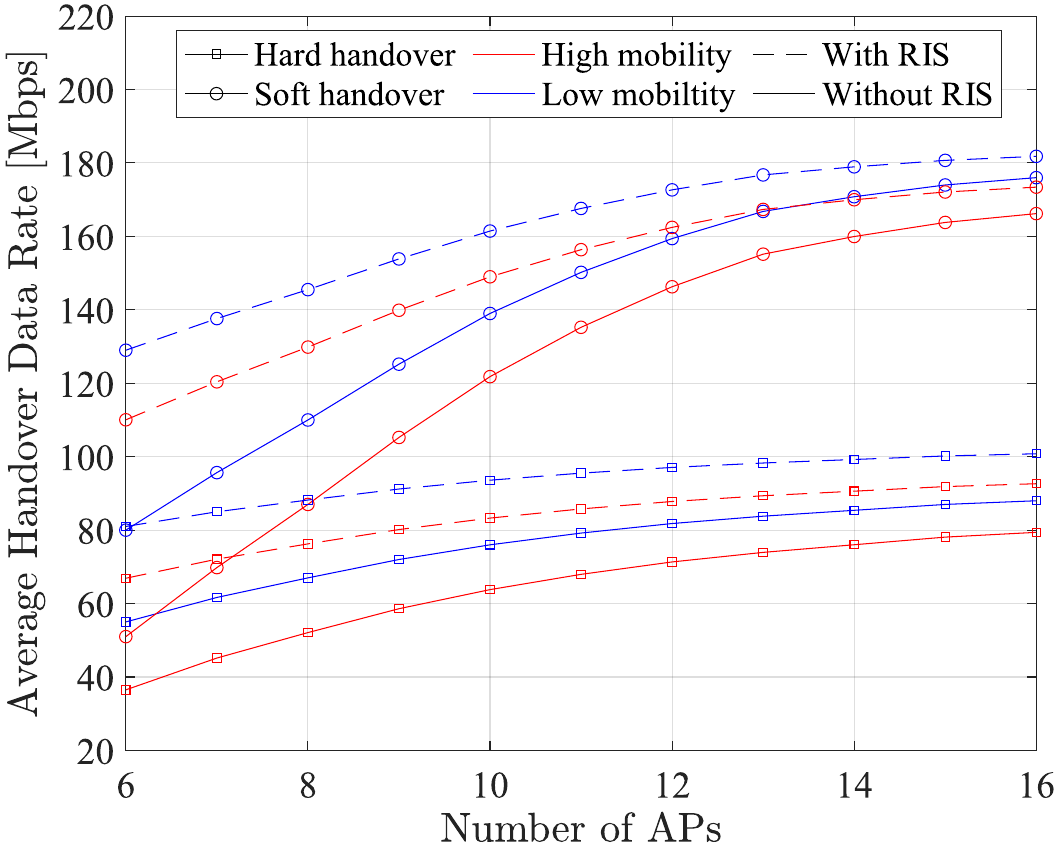}
        \vspace{-2mm}
    \caption{Average handover data rate vs. number of APs.}
    \label{fig:result1}
    \end{center}
    \vspace{-6mm}
    \end{figure}

\figurename~\ref{fig:result1} shows the average handover data rate versus the number of APs in the cases of low mobility, high mobility, hard handover, and soft handover. Moreover, results are compared with the case of no RISs being involved. Our results verify that the use of RIS results in data rate improvement in almost all cases. Especially, in the case of a small number of APs, the percentage improvement of the data rate is comparably high (\textit{i.e.}, upto $60\%$) compared to the case of a large number of APs. This is expected since RIS avoids connectivity holes that can be present when a low number of APs are used. Both soft and hard handovers show a saturating nature in the high APs regime, due to zero connectivity holes. In general, the soft handover results in better performance than the hard handover, and especially, up to $80\%$ data rate improvement can be observed. The performance improvement with the RIS in the soft handover is higher than in the hard handover scenario as a result of focusing light from several APs to the receiver using the ANN method. 

    \begin{figure}[t]
    \begin{center}
    %\vspace{-1mm}
    \includegraphics[width=0.78\columnwidth]{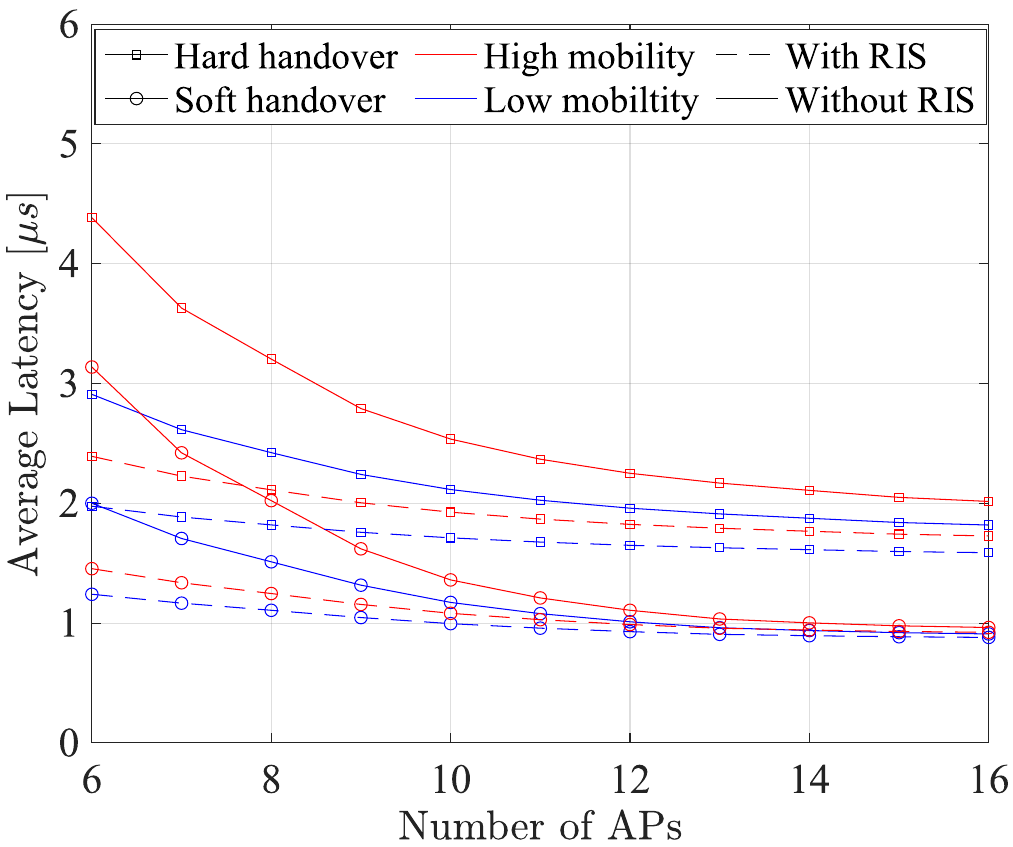}
    %\vspace{-1mm}
            \vspace{-2mm}
    \caption{Average handover latency vs. number of APs.}
    \label{fig:result2}
    \end{center}
    \vspace{-6.5mm}
    \end{figure}
    
\figurename~\ref{fig:result2} shows the average handover latency versus the number of APs. The handover latency shows a decreasing nature when the number of APs is increased which is a counter-trend to the data rate since it is inversely proportional to the data rate. The latency saturates around $1 \mu s$ and $2 \mu s$ at the cases of soft and hard handover cases. The use of RIS guarantees a low latency compared to the case without RISs. Especially, latency is almost constant for different numbers of APs when RIS is used. For the considered setup, the handover latency lies between $1 \mu s$ and $4.5 \mu s$ for all the cases. Though it is illustrated, the use of RIS results in a decrease in the number of handovers in all cases compared to no RIS scenario. The handover decrease is high for soft handover compared to hard handover. Specifically up to $12\%$ handover reduction is expected in soft handover case compared to hard handover.

%\vspace{-1mm}

\section{Conclusions}
In this paper, we presented an RIS-assisted proactive handover mechanism for indoor VLC systems under blockage conditions. Especially, under blockage conditions induced by human movements, the use of mirror-based RIS is a promising approach to improving the handover performance. We present an algorithm to perform the hard and soft handover and RIS assignment for the considered scenario and is supported by an ANN architecture. Our results show that the use of RIS improves the handover performance and helps to achieve high data rates and low handover latency. Specifically, data rate improvement up to $60\%$ is expected. The ANN-based RIS assignment results in a very low amount of connectivity holes and guarantees
a smooth handover. It is noted that soft handover results in better performance compared to hard handover in most cases. Specifically, upto $12\%$ of handovers can be reduced using soft handover compared to hard handover for high mobility. 

\vspace{-2mm}
\section*{Acknowledgment}
This article is partially based on work from COST Action NEWFOCUS CA19111, supported by COST (European Cooperation in Science and Technology) and has received funding from the European Research Council (ERC) under the European Union's Horizon 2020 research and innovation programme (Grant agreement No. 819819).

\bibliographystyle{IEEEtran}
\bibliography{Main}  
 
\end{document}